\documentclass[dvips]{article}
\usepackage{icrctc07}

\title{Future GLAST observations of Supernova remnants and Pulsar Wind
Nebulae}
\shorttitle{GLAST observations of SNRs and PWNe}

\authors{S. Funk$^{1}$ for the GLAST collaboration}
\shortauthors{Funk et al.}
\afiliations{
  $^1$Kavli Institute for Particle Astrophysics and
  Cosmology, SLAC, Menlo Park, CA-94025, USA\\ 
}

\email{Stefan.Funk@slac.stanford.edu}

\abstract{Shell-type Supernova remnants (SNRs) have long been known to
harbour a population of ultra-relativistic particles, accelerated in
the Supernova shock wave by the mechanism of diffusive shock
acceleration. Experimental evidence for the existence of electrons up
to energies of ~100 TeV was first provided by the detection of hard
X-ray synchrotron emission as e.g. in the shell of the young SNR
SN1006. Furthermore using theoretical arguments shell-type Supernova
remnants have long been considered as the main accelerator of protons
- Cosmic rays - in the Galaxy; definite proof of this process is
however still missing.  Pulsar Wind Nebulae (PWN) - diffuse structures
surrounding young pulsars - are another class of objects known to be a
site of particle acceleration in the Galaxy, again through the
detection of hard synchrotron X-rays such as in the Crab
Nebula. Gamma-rays above 100 MeV provide a direct access to
acceleration processes. The GLAST Large Area telescope (LAT) will be
operating in the energy range between 30 MeV and 300 GeV and will
provide excellent sensitivity, angular and energy resolution in a
previously rather poorly explored energy band. We will describe
prospects for the investigation of these Galactic particle
accelerators with GLAST.}

\begin{document}
\maketitle
\section{Shell-type Supernova remnants}
Supernova remnants, through shocks in their expanding shells, have
long been thought to accelerate charged particles to
ultra-relativistic energies~\cite{Ginzburg, Roger}. These charged
particles can subsequently emit radio, X-rays or gamma-rays through
interactions with magnetic fields and surrounding material. In spite
of recent detailed studies of Supernova remnants in particular with
VHE gamma-rays~\cite{HESS1713, HESSVelaJr}, the nature of the parent
population responsible for the gamma-ray emission remains elusive. It
is not yet evident, whether the bulk of the gamma-rays are produced by
Bremsstrahlung or Inverse Compton (IC) scattering of electrons, or by
hadronic interactions and subsequent $\pi^0$-decay. If in the future a
hadronic origin of the gamma-ray emission can be established, this
would represent a great step towards the final proof that shell-type
SNRs are the long sought source of cosmic rays in the Galaxy. In the
GeV band EGRET data showed a statistical associations of gamma-ray
emission with radio SNRs (and related sources)~\cite{EGRETSNR},
however, no individual shell-type SNR could unambiguously be
identified. The upcoming GLAST-LAT instrument, however, has the
spectral and angular resolution to perform first detailed study of
these object between 30~MeV and 300~GeV.
\subsection{Spectral Studies of shell-type Supernova remnants}
\begin{figure}
\begin{center}
\noindent
\fbox{\hbox{\vbox{\hsize=45mm 
      \includegraphics [width=53mm]{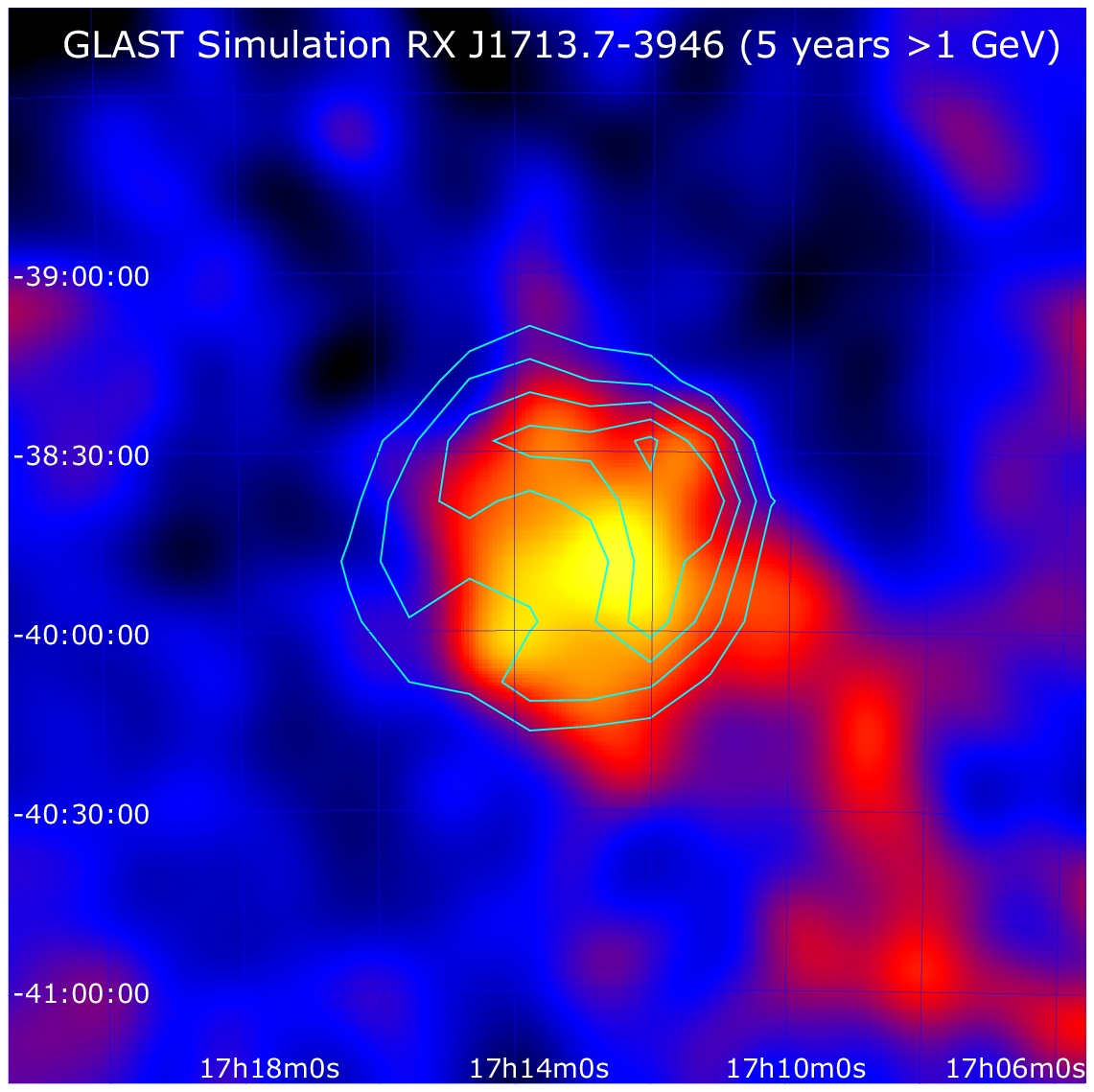}
      \includegraphics [width=45mm]{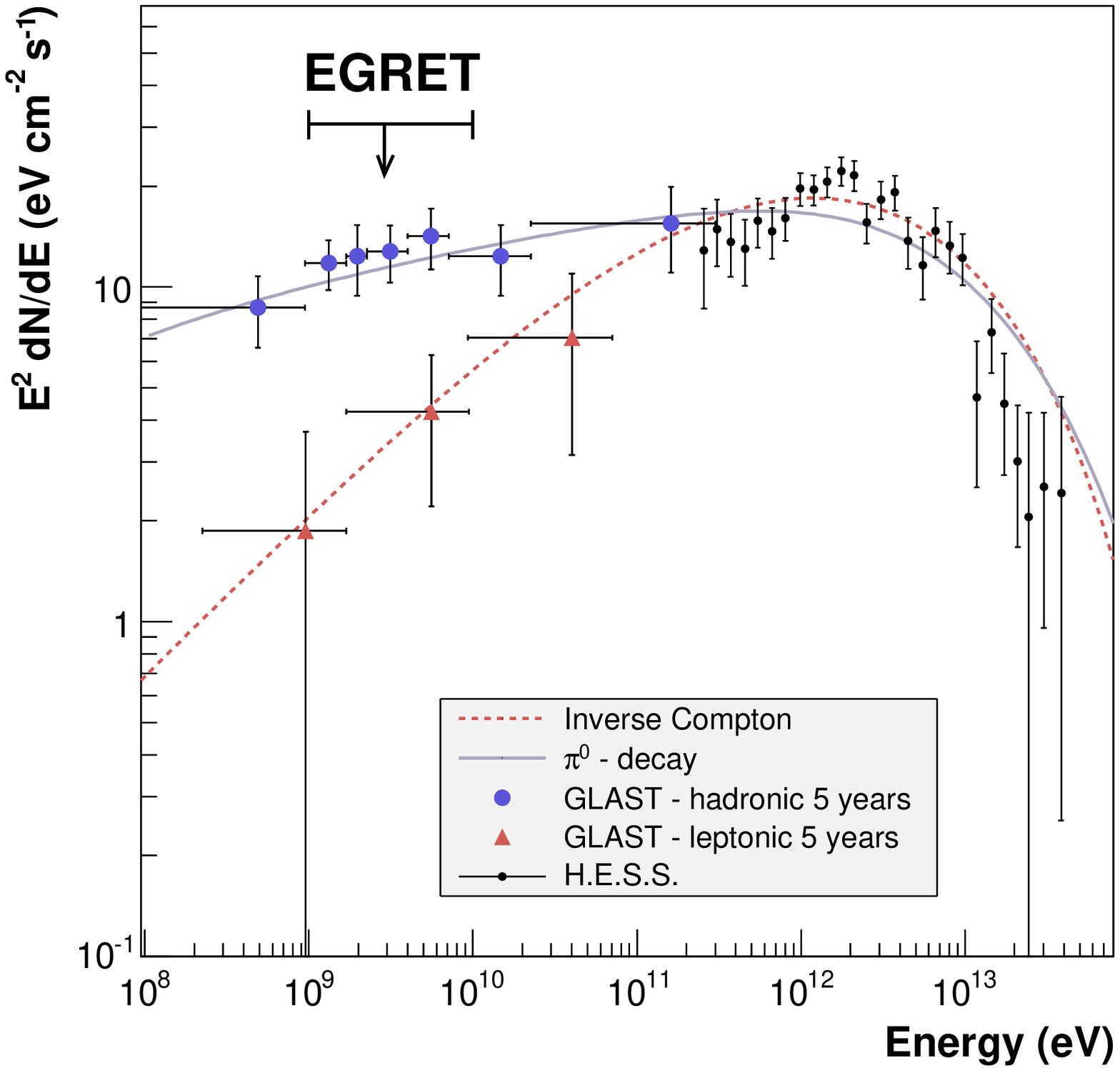}
}}}
\end{center}
\caption{ {\bf{Top:}} Simulated map (smoothed counts) for
RX\,J1713.7--3946 in a hadronic model above 1~GeV as seen by the LAT
in 5 yrs of observation (H.E.S.S.\ contours in blue).  {\bf{Bottom:}}
SED for the same SNR. H.E.S.S. data are shown in black, simulated
GLAST-LAT spectra are shown for different emission mechanisms of the
gamma-rays (blue: hadronic, red: leptonic).}\label{fig::1713}
\end{figure}

The GLAST-LAT will provide measurements of gamma-ray spectra between
30 MeV and 300 GeV, a previously rather poorly-explored energy
regime. LAT data will allow us to distinguish between different models
for the gamma-ray emission. Gamma-rays of leptonic origin (produced by
IC) can in principle be distinguished from those of hadronic origin
(produced by $\pi^0$-decay) through their characteristic spectral
shape, although recent claims have been made that under certain
conditions the leptonic gamma-ray spectra might resemble those of
pionic decays~\cite{Ellison}.  Previous measurements in higher energy
gamma-rays above 100 GeV provide rather stringent constraints on the
absolute emission level for the different
models. Figure~\ref{fig::1713} (bottom) shows predictions for
GLAST-LAT measured energy spectra (in 5 years of scanning
observations) for a hadronic and a leptonic emission scenario
illustrating that the LAT energy range is particularly well suited to
distinguish these models and potentially provide the first direct
evidence of hadronic acceleration in the shells of SNRs.

\subsection{Morphological Studies of shell-type Supernova remnants}

The unprecedented GLAST-LAT angular resolution will alleviate the
problem of source confusion in the Galactic plane and will allow for
studies of the gamma-ray emission regions in the larger of the known
gamma-ray emitting SNRs, such as RX\,J0852.0--4622 (also known as Vela
Junior). The angular resolution of the instruments follows the
relation $\delta \Theta = 0.6\mathrm{(E/GeV)}^{-0.8}$, resulting in an
angular resolution of $\sim 0.6^{\circ}$ at 1 GeV. Young nearby SNRs
with angular size in excess of $\sim 1^{\circ}$ can thus be
significantly resolved above $\sim 3 - 10$ GeV provided sufficient
photon flux at these high energies. The brightest VHE gamma-ray SNR
RX\,J1713.7--3946 is shown in Figure~\ref{fig::1713} (top) above an
energy of 1~GeV for 5 years of simulated data in scanning mode. This
object will be barely resolvable unless deconvolution methods are
applied. Correlation studies of GeV SNR candidates with hard X-rays,
as well as with VHE gamma-rays will give detailed views into the
acceleration sites, providing an energetic coverage of many orders of
magnitude. The excellent angular resolution will isolate the
shell-emission from the core PWN emission in nearby composite SNRs and
allow for population studies of shell-type SNRs in the gamma-ray
regime.

\section{GLAST studies of Pulsar Wind Nebulae}
\begin{figure}
\begin{center}
\noindent
\fbox{\hbox{\vbox{\hsize=60mm 
      \includegraphics
      [width=53mm]{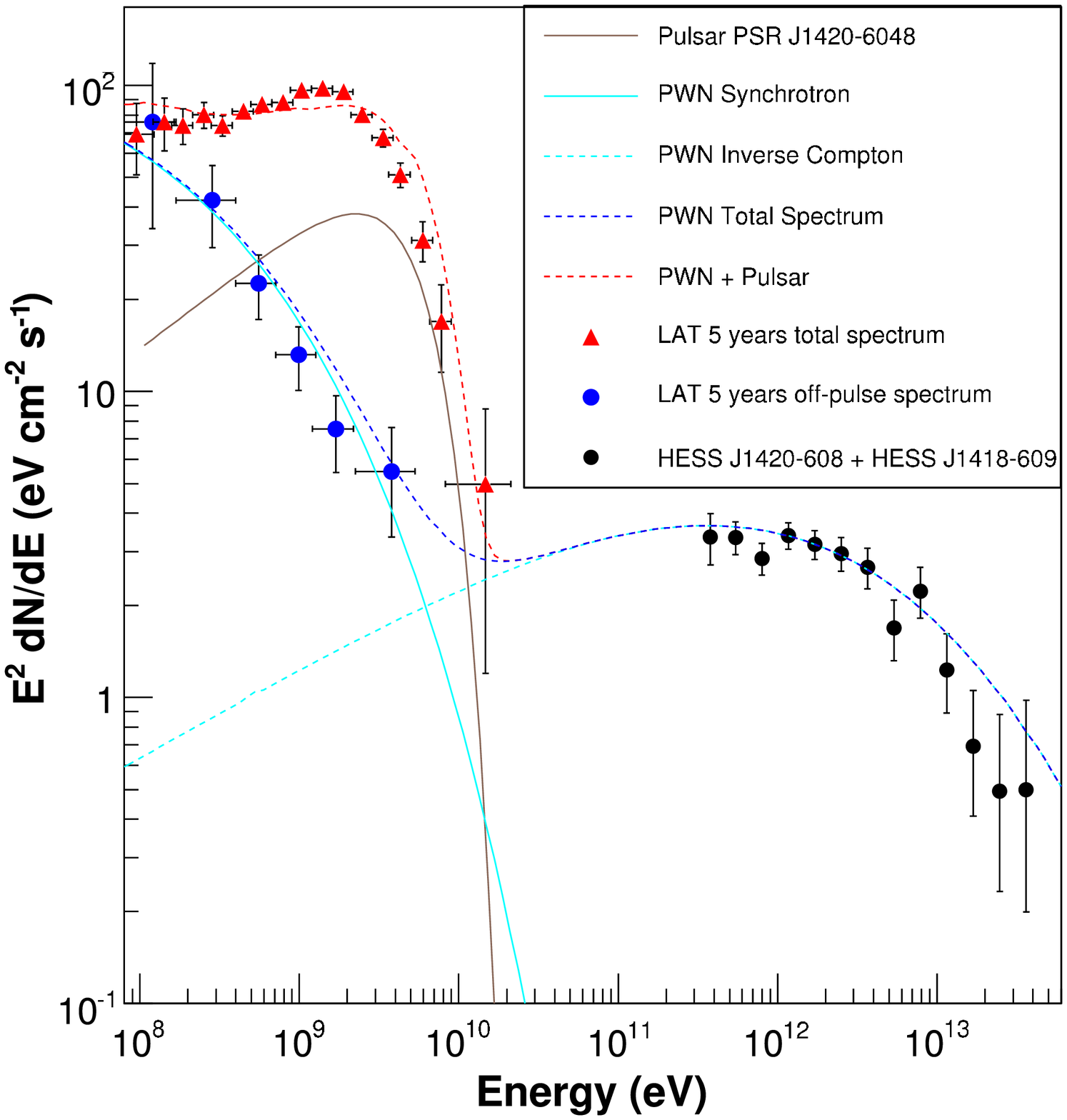} 
      \includegraphics
      [width=53mm]{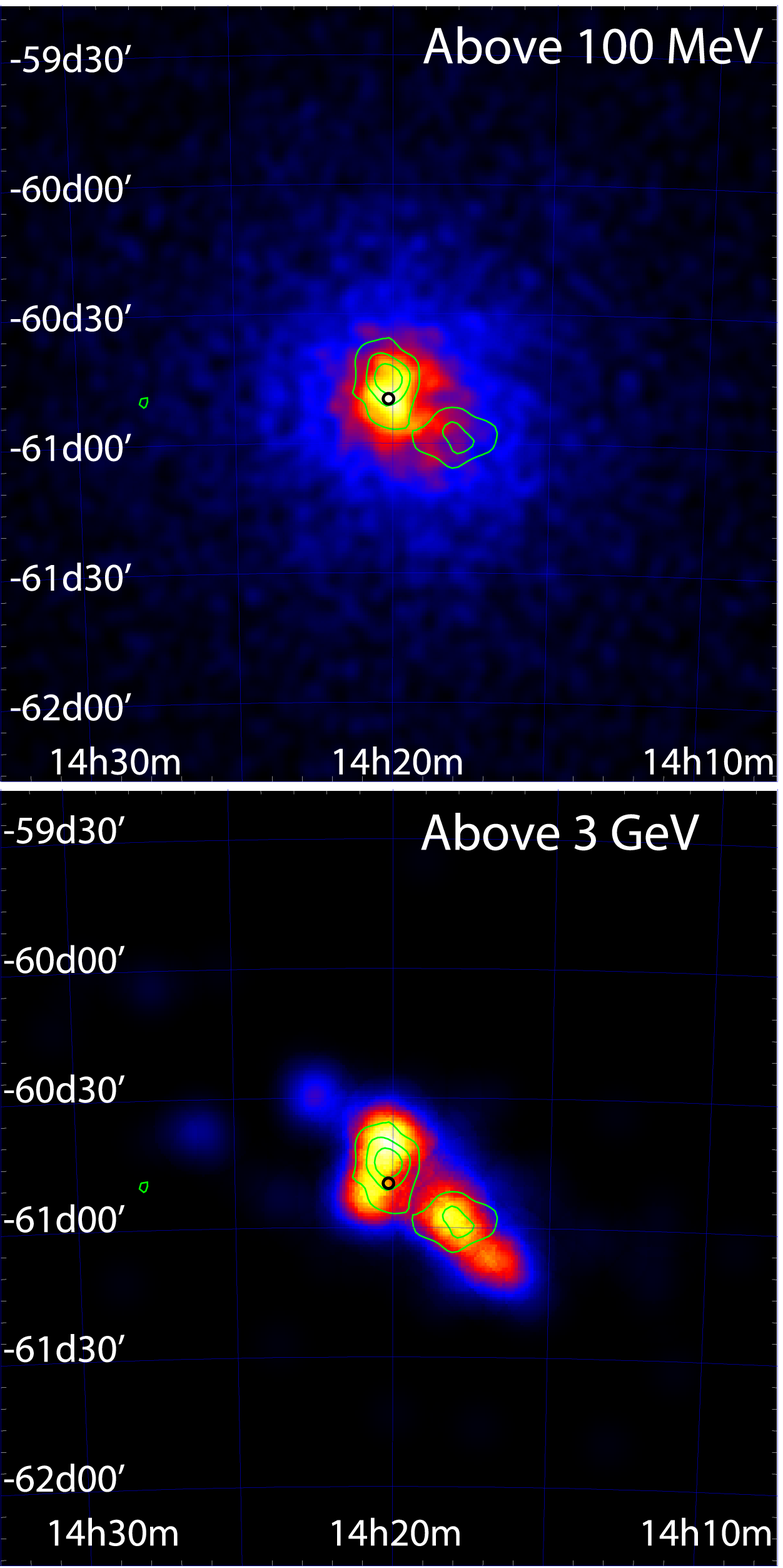}
}}}
\end{center}
\caption{{\bf{Top:}} Spectral Energy distribution for the Kookaburra
region, containing a simulated Pulsar and PWN in the GLAST range and
the VHE gamma-ray measurements by
H.E.S.S.~\cite{HESSKooka}. {\bf{Center:}} Corresponding GLAST
gamma-ray image simulated as above for energies $>100$~MeV, and
{\bf{Bottom:}} $>3$~GeV).}\label{fig::Kooka}
\end{figure}

EGRET found a number of bright variable Galactic objects that are
potentially associated with Pulsar Wind Nebulae (PWN). Recent advances
in VHE gamma-rays above 100 GeV by H.E.S.S. have shown that there are
at least 8 PWN emitting at gamma-ray energies detected in a survey of
the southern Galactic plane~\cite{HESSSurvey}. Depending on the
position of the peak of the IC emission, several of these are expected
to be visible in the GLAST band, in particular because GLAST can probe
large angular scales, generally not easily accessible in lower energy
bands like radio or X-rays (as shown by H.E.S.S.\ in cases such as
HESS\,J1825--137 and HESS\,J1640--465 the high-energy IC gamma-ray PWN
can show a larger extent than the X-ray PWN, a property that makes
gamma-ray instruments with their wide fields of view ideal instruments
to detect these). Figure~\ref{fig::Kooka} shows a simulation of GLAST
data for the Kookaburra region. The spectral energy distribution shown
in the top panel demonstrates that the GeV emission should be
dominated by the central pulsar in this region. However, phase
analysis can cut out the pulsed emission, revealing the $> 100$ MeV
PWN spectrum. This is also illustrated in the lower figures, showing
the GLAST simulated 2D-map above 2 different energies. The upper plot
above 100 MeV is completely dominated by the pulsed photons, the lower
panel above 3 GeV allow morphological studies of the region, due to
the strong cutoff in the pulsar spectrum. GLAST will be able to
determine morphologies and energy spectra for a number of PWN and
allow for population studies. Because of the near continuous coverage
and stable high sensitivity of GLAST, it is expected that slow
(month-year) variability of the PWN synchrotron component from the
wind termination shock should be measurable in some cases providing a
new probe of PWN dynamics. Recently Reimer\&Funk have shown that by
using the high angular resolution gamma-ray detection as provided by
H.E.S.S.\, the EGRET data on the Kookaburra region can be disentangled
into two distinct sources~\cite{Reimer}. This might be a template case
for future studies of such systems in crowded regions in the Galactic
plane where GLAST will have to fight with source confusion.

\section{Best candidates for GLAST detections}
\paragraph{Shell-type SNRs}: 
The best candidates for detecting gamma-ray emission in the GLAST-LAT
energy range are a) SNRs that have been detected in VHE gamma-rays
such as RX\,J1713.7--3946 or b) other young SNRs that emit hard x-ray
(synchrotron emission) in their shells or c) older SNRs like e.g. W28
which had sufficient time to accumulate a large amount of cosmic rays
in their shells and which potentially have dense molecular clouds in
their vincinities that could act as target material for hadronic
interactions with subsequent pionic decays into gamma-rays. The best
candidates are summarised in table~\ref{tab::candidates}.

\begin{table}[t]
\begin{center}
\begin{tabular}{c|c}
\hline
Candidate SNR & Candidate PWN  \\
\hline
RX\,J1713.7--3946 & Crab Nebula\\
RX\,J0852.0--4622 & Vela~X\\
Cas~A & Kookaburra\\
SN\,1006 & MSH\,15-52\\
RCW\,86 & PSR\,B1706--44\\
W\,28 & PSR\,B1823--13\\
Tycho~SNR & 3C58\\
Kepler SNR & MSH\,11-54\\
IC~443 & CTB~80\\
\hline
\label{tab::candidates}
\end{tabular}
 \caption{Candidate SNRs and PWNe that might be detectable with the
 GLAST-LAT.}

\end{center}
\end{table}

\begin{figure*}
\begin{center}
\noindent 
\fbox{\hbox{\vbox{\hsize=140mm 
\includegraphics
    [width=140mm]{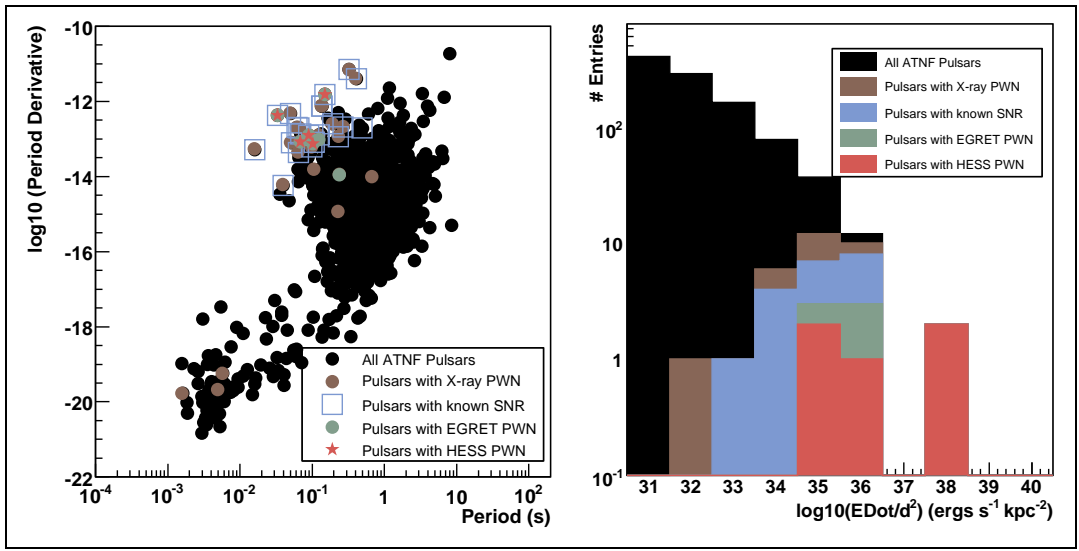}}}}
\end{center}
\caption{{\bf{Top:}} $\mathrm{P}-\dot{\mathrm{P}}$ diagram for
pulsars: all ATNF pulsars (black), with detected X-ray PWN (brown),
with a known corresponding SNR (blue), potentially associated to an
EGRET source (green), associated to a H.E.S.S.\ VHE PWN
(red). {\bf{Bottom:}} Energy output for the selections used at the
top.}\label{fig::Pulsars}
\end{figure*}

\paragraph{Pulsar Wind Nebulae}: 
The best candidates for gamma-ray PWN are a) PWN detected in VHE
gamma-rays b) PWN detected in X-rays. About 30 X-ray PWN have been
detected mostly around young energetic pulsars as shown in
Figure~\ref{fig::Pulsars}. A significant correlation has now been
established between VHE gamma-ray sources above 100~GeV and energetic
pulsars (in terms of {$\dot{E}/d^2$}) from the ATNF catalogue (see
Carrigan, these proceedings). Even though the GLAST gamma-ray fluxes
might be contaminated by photons from the pulsar, the most energetic
pulsars are a promising target for the detection of a PWNe surrounding
the central pulsar (see Table~\ref{tab::candidates}).

\section{Summary} The prospects for GLAST of detailed investigations
of SNRs and PWNe promise to provide a sensitive new probe of particle
acceleration mechanisms in our Galaxy . Measurements in adjacent X-ray
and VHE $\gamma$-ray energy bands allow for detailed predictions of
possible $\gamma$-ray signatures in the GLAST energy range.

\bibliographystyle{plain}

\begin{thebibliography}{}
\bibitem{Ginzburg} Ginzburg, V.~L., \& Syrovatskii, S.~I. 1964
\bibitem{Roger} Blandford, R.~D., \& Ostriker, J.~P., 1978, ApJ 221,
L29
\bibitem{HESS1713} Aharonian, F., et al., 2007, A\&A 464, 235
\bibitem{HESSVelaJr} Aharonian, F., et al., 2007, ApJ 661, 236
\bibitem{EGRETSNR} Sturner, S.~J., \& Dermer, C.~D., 1995, A\&A 293
(1), L17
\bibitem{Ellison} Ellison, D.~C., et al, 2007, submitted to ApJ
  (astro-ph/0702674)
\bibitem{HESSSurvey} Aharonian, F., et al., 2006, ApJ 636, 777
\bibitem{HESSKooka} Aharonian, F., et al., 2006, A\&A 456, 245
\bibitem{Reimer} Reimer, O., \& Funk, S., 2006, in press, Ap\&SS
  (astro-ph/0611653)
\end{thebibliography}

\end{document}